\documentclass[12pt,epsfig,rotate]{article}
\usepackage{epsfig}
\usepackage{wrapfig}
\usepackage{graphicx}
\begin{document}

\setcounter{page}{1}

\begin{center}
{\large\bf Polarization parameters
 of the quasi--elastic $(p,2p)$ reaction
with nuclei at 1~GeV }\\
\vspace*{9mm}
O.V.~Miklukho, G.M.~Amalsky, V.A.~Andreev,
S.V.~Evstiukhin, O.Ya.~Fedorov, G.E.~Gavrilov, A.A.~Izotov, A.Yu.~Kisselev, 
L.M.~Kochenda, M.P.~Levchenko, V.A.~Murzin, D.V.~Novinsky,
A.N.~Prokofiev, A.V.~Shvedchikov, S.I.~Trush, A.A.~Zhdanov\\
\vspace*{5mm}
{\it  B.P.~Konstantinov Petersburg Nuclear Physics Institute,
Gatchina, RUSSIA}
\end{center}
\vspace*{10mm}
New experimental data on the polarization and spin-correlation parameters
in the {\it (p,2p)} reaction with nuclei at 1 GeV are presented. The experiment
was aimed to study a modification of the proton--proton scattering matrix.\\
\\
\\
\\
{\bf Comments:} 11~pages, 5~figures. Presented at XV International Workshop 
on High--Energy Spin Physics, Dubna, October 2013. Submitted to the Workshop proceedings.\\
\\
\\
\\
{\bf Category:} Nuclear Experiment (nucl--ex)\\
\newpage
\section{Introduction}
~~~~There were some speculations on  modifications of nucleon and meson masses
and sizes, and of meson--nucleon coupling constants, and, as a
consequence, of nucleon--nucleon scattering matrix in nuclear medium [1--3].
These speculations were
motivated by a variety of theoretical points of view, including the
renormalization effects due to strong relativistic nuclear fields,
deconfinement of quarks, and partial chiral symmetry restoration.

 This  work is a part of the experimental program in the framework of which
 the medium--induced modifications of the nucleon--nucleon scattering amplitudes are
 studied at the PNPI synchrocyclotron with the 1~GeV proton beam [4--8].
The intermediate--energy quasi--free $(p,2p)$ reaction is a good experimental
tool to study such effects, since in the first approximation, this reaction can
be considered as a proton--proton scattering in the nuclear matter. Usage of
S--shell protons (with zero orbital momentum) is preferred because interpretation
of obtained data in this case is essentially simplified since the
effective polarization is not involved [9].
The polarization observables in the reaction are compared with
those in the  elastic $pp$ scattering.
In our exclusive experiment, a two--arm magnetic spectrometer
 is used, the shell structure of
 the nuclei being evidently distinguished. To measure polarization characteristics of the
 reaction, each arm of the spectrometer was equipped with a multi-wire--proportional chamber
 polarimeter.

 In the early PNPI--RCNP experiment [5], the polarizations $P_1$  and $P_2$ of both
 secondary protons from the $(p,2p)$
 reactions at 1~GeV with the 1$S$--shell protons of the nuclei $^6$Li, $^{12}$C and
 with the 2$S$--shell protons of the $^{40}$Ca nucleus has been measured at
 nuclear proton momenta
 close to zero. The polarization observed in the experiment,
 as well as the analyzing power $A_y$ in the RCNP experiment
at the 392~MeV polarized proton beam [10,~11],
 drastically differed from those calculated in the framework of non--relativistic Plane Wave Impulse
 Approximation (PWIA) and of spin--dependent Distorted Wave Impulse Approximation (DWIA) [12],
based on free space proton--proton interaction. This difference was found to have a negative value
and to increase monotonously with the effective mean nuclear density
$\bar\rho$ [10]. The latter  is determined by the
 absorption of initial and secondary protons in nucleus matter.
 The observed inessential difference between the non--relativistic PWIA and DWIA
 calculations pointed out only to a small
depolarization of the secondary protons because of  proton--nucleon re-scattering inside a nucleus.
 All these facts strongly indicated
a modification of the proton--proton scattering amplitudes due to the modification of the main properties
of hadrons in the nuclear matter.

 Later, the results of the experiment with a $^4$He target
broke the above--mentioned dependence of the difference
between the experimental polarization values  and those
calculated in the framework of the PWIA on the effective mean nuclear
density [6].
The difference for the $^4$He nucleus proved
to be smaller than that for the $^{12}$C nucleus. This evidently contradicts the elastic
proton--nucleus scattering experiment. According to the experiment, the $^4$He nucleus
has the largest mean nuclear density.
The important feature of the experiment with the $^4$He nucleus
was a possibility to see the medium effect without any contribution from multi--step processes
(for instance, from the  $(p,2pN)$ reactions). These processes  could take place when
there were nucleons of outer shells as in other nuclei. Therefore, they
could not cause the systematic difference between the polarizations
$P_1$ and $P_2$  clearly obtained for the first time in the experiment [6].

Here we present the polarization data for the reaction with the nuclei
$^4$He, $^6$Li, $^{12}$C (1$S$--shell), and $^{40}$Ca (2$S$--shell) obtained
with a much better statistical accuracy in our last experiments.
New data on the polarization
in the reaction with the 1$S$--shell protons
of the $^{28}$Si nucleus are presented. The 1$S$--state of the  $^{28}$Si nucleus has
a larger value of the mean proton binding energy $E_s$ (50~MeV) than  that of the $^{12}$C nucleus
(35~MeV). We also present the polarization measured in
the reaction with the $P$--shell and $D$--shell protons of the $^{12}$C and $^{28}$Si
nuclei, respectively.

In recent experiments, the research program was extended to measure the spin
correlation parameters $C_{ij}$  in the $(p,2p)$ reaction with the $^4$He and $^{12}$C
nuclei. Measurements of the parameters in the reaction with nuclei were for the first
time performed.
The main attention was concentrated on the spin correlation
parameter $C_{nn}$  since its value is the same in the
center--of--mass and laboratory systems. Besides, this parameter is not distorted by
the magnetic fields of the two--arm spectrometer because of the proton
anomalous magnetic moment [13]. Since the polarization  and  the
spin correlation parameter $C_{nn}$ are expressed differently  through the
scattering matrix elements [3], the measurement of both these
polarization observables  can
provide a more comprehensive information about a modification of the hadron
properties in the nuclear medium.
\section{Experimental method}
~~~ The general layout of the
experimental setup is shown in Fig.~1 [14].
The experiment is performed at non--symmetric scattering angles
of the final state protons in the coplanar quasi--free scattering
geometry with a complete reconstruction of the reaction
kinematics.
The measured secondary proton momenta $K_1$, $K_2$ (kinetic energies $T_1$, $T_2$)
and the scattering angles $\Theta_1$, $\Theta_2$ are used together
with the proton beam energy $T_0$ to calculate nuclear proton
separation energy $\Delta E$~=~$T_0$-$T_1$-$T_2$ and the residual nucleus momentum {\bf $K_r$}
for each $(p,2p)$ event. In the impulse approximation, the $K_r$
is equal to the momentum $K$ of the nuclear proton before the
interaction (${\bf K_r}$~=~-${\bf K}$).

 External proton beam of the PNPI synchrocyclotron was focused
onto the target TS of a two--arm spectrometer consisting of the magnetic
spectrometers MAP and NES. The beam intensity was monitored by
the scintillation telescope M1, M2, M3 and was at the level of about
5$\cdot$10$^{10}$~protons/(s$\cdot$cm$^2$).

Solid nuclear targets TS made of CH$_2$ (for the setup
calibration), $^6$Li, $^{12}$C, $^{28}$Si, and $^{40}$Ca, as well as
a cryogenic target made of liquid helium  $^4$He (or
liquid hydrogen for calibration) were used in the
experiment [6,~14].

The spectrometers were used for registration of the
secondary protons from the $(p,2p)$ reaction in coincidence and for
measurement of their momenta and outgoing angles. The polarization
of these protons $P_1$ and $P_2$, 
\begin{figure}
\centering\epsfig{file=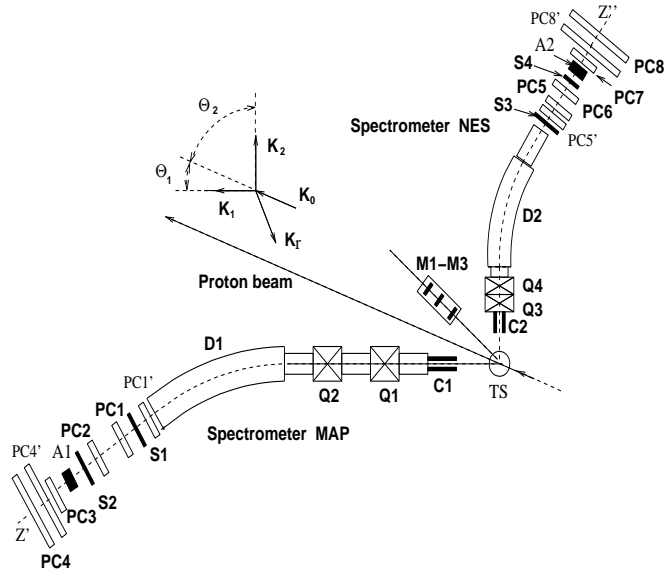,width=.6\textwidth,height=75mm}
\caption{\small The experimental setup. TS is the target of the two--arm spectrometer;
Q1$\div$Q4 are magnetic quadrupoles; D1, D2 are  dipole
magnets; C1, C2 are  collimators; S1$\div$S4 and M1$\div$M3 are
 scintillation counters; PC1$\div$PC4, PC1', PC4'
(PC5$\div$PC8, PC5', PC8') and A1 (A2) are the proportional
chambers and the carbon analyzer of the high--momentum (low--momentum)
polarimeter, respectively.}
\end{figure}
and the spin correlation
parameters $C_{ij}$ were measured by the polarimeters located in
the region of focal planes of the spectrometers MAP and NES (Fig.~1).
The first index of the $C_{ij}$,
$i$ (where $i$ is $n$ or $s^,$), and the second index $j$ (where $j$ is $n$ or $s^{,,}$)
correspond to the forward scattered proton analyzed by the
MAP polarimeter and the recoil proton analyzed by the NES
polarimeter, respectively. The unit vector ${\bf n}$ is
perpendicular to the  scattering plane of the  reaction.
Unit vectors ${\bf {s^,}}$ and ${\bf s^{,,}}$
are perpendicular to the vector ${\bf n}$ and to the coordinate axes
z$^,$ and z$^{,,}$ (Fig.~1) of the polarimeters.

The overall energy resolution (on $\Delta E$) of the spectrometer estimated from the elastic
proton--proton scattering with the 22~mm thick cylindrical CH$_2$
target was about 5~MeV (FWHM). The spectra which was analysed is presented in Fig.2 [14].
\begin{figure}
\centering\epsfig{file=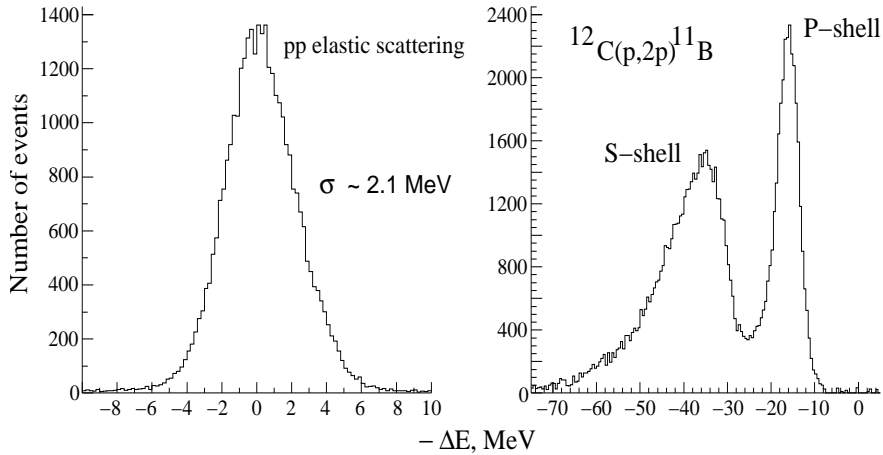, width=.8\textwidth,height=60mm}
\caption{\small Proton separation energy spectra for elastic $pp$--scattering (left panel) 
and for the $(p,2p)$ reaction with $^{12}$C nuclei (right panel). In the $^{12}$C spectrum the 
accidental background contribution is subtracted.}
\end{figure}

The track information from the proportional chambers of both
polarimeters was used in the off-line analysis to find the
azimuthal $\phi_1$, $\phi_2$ and polar $\theta_1$, $\theta_2$
angles of the proton scattering from the analyzers A1, A2 for each
$(p,2p)$ event.

The polarization parameters were estimated by
folding the theoretical functional shape of the azimuthal angular distribution
into experimental one [8], using the CERNLIB MINUIT
package  and a  $\chi^2$ likelihood estimator. This method
permits to realize the control over $\chi^2$ in the case the
experimentally measured azimuthal distribution is distorted due to
the instrumental problems.

The time difference (TOF) between the signals from the
scintillation counters S2 and S4 was measured. It
served to  control  the accidental coincidence background. The
events from four neighboring proton beam bunches were recorded.
Three of them contained the background events only and were used in
the off-line analysis to estimate the background polarization
parameters and the background contribution at the main bunch containing
the correlated events [14].

  The recoil proton spectrometer NES  was installed at
a fixed angle $\Theta_2\simeq$ 53.2$^\circ$.
At a given value of the $S$--shell mean binding energy of the nucleus
under investigation, the angular and momentum settings of the MAP
spectrometer and the momentum setting of the NES spectrometer were
chosen to get a kinematics of the $(p,2p)$ reaction close to that of the
elastic proton--proton scattering. In this kinematics,
the momentum  $K$ of the nuclear $S$--proton before the interaction is
close to zero. At this condition, the counting rate of the $S$-shell proton
knockout reaction should be maximal.

\section{Results and discussion}
~~~ In Fig.~3, the polarizations $P_1$, $P_2$  in the $(p,2p)$ reaction
with the $S$--shell protons of the nuclei $^4$He, $^6$Li, $^{12}$C, $^{28}$Si, $^{40}$Ca
are plotted versus  the $S$--shell proton binding energy $E_s$ [14].
For all nuclei (excluding $^4$He),
the effective mean nuclear density $\bar\rho$, normalized on the saturation nuclear
density $\rho_0\approx$~0.18 fm$^{-3}$, is given.
 The actual calculation of the effective mean
nuclear density $\bar\rho$, which is determined by absorption of the incident
and both outgoing protons, was carried out following a procedure
[10] using the computer code THREEDEE [12].
The potential model of a nucleus employed by the code is not  correct for
the $^4$He nucleus. The calculated value of the $\bar\rho$ in this case
is strongly unreliable [6]. The $^4$He data should be excluded
in comparison with theoretical models which differ from the PWIA.

The points
($\circ$) and ($\bullet$) in the figure correspond to the
polarizations $P_1$
\begin{figure}
\centering\epsfig{file=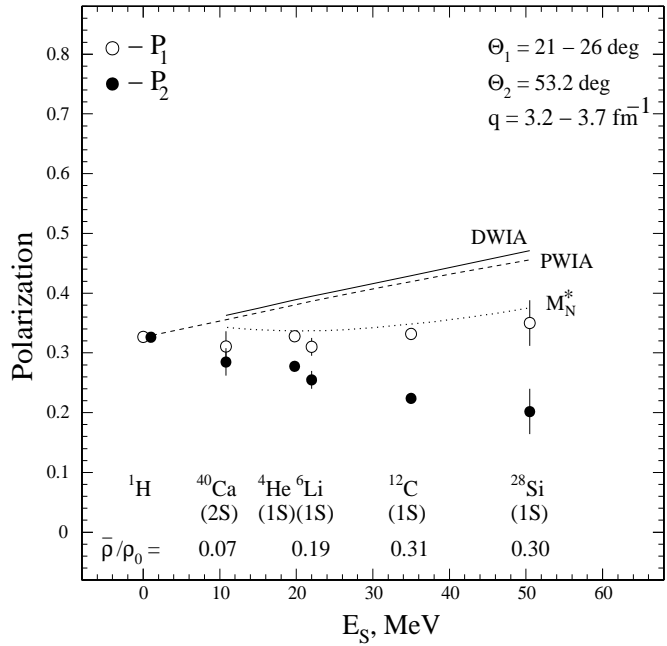,width=.6\textwidth,height=85mm}
\caption{\small Polarizations $P_1$ and $P_2$  of the  protons scattered
at the angles $\Theta_1$  ($\circ$) and $\Theta_2$ ($\bullet$)
in the $(p,2p)$ reaction with the $S$--shell protons
of nuclei at 1~GeV. The points at
$E_s$~=~0 correspond to the elastic proton-proton scattering.
The curves correspond the theoretical calculations
described in the text.}
\end{figure}
and $P_2$ of the forward scattered protons at the
angle $\Theta_1$~=~21$^\circ\div$25$^\circ$ (with
energy $T_1$~=~745$\div$735~MeV) and of the recoil protons
scattered at the angle $\Theta_2\simeq$ 53.2$^\circ$ (with
energy $T_2$~=~205$\div$255~MeV). The points at the $E_s$=0
are the polarizations $P_1$ and $P_2$
in the elastic proton-proton scattering at the angles $\Theta_1$~=~26.0$^\circ$
and $\Theta_2$~=~53.2$^\circ$ ($\Theta_{cm}$~=~62.25$^\circ$).

In Fig.~3, the experimental data are compared with the results of the
non-relativistic PWIA, DWIA calculations (the dashed and solid curves, respectively) [14]
and the DWIA* calculation with the relativistic effect, the distortion of
the nucleon Dirac spinor in nuclear medium,  taken into account (the dotted, $M^*_N$,
curve) [2,14].
For the calculations, the computer code THREEDEE was used [12]
using
an on--shell factorized approximation and the final energy prescription.
A global optical potential,
parametrized in the relativistic framework and converted to the
Shr\"{o}dinger--equivalent form, was used to calculate the
distorted wave functions of incident and outgoing protons in the case of
DWIA and DWIA*. A conventional well--depth method was used to construct
the bound--state wave function. The DWIA* calculations were
carried out in the Shr\"{o}dinger--equivalent form [5]. In
this approach, a modified $NN$ interaction in medium is
assumed due to the effective nucleon mass (smaller than the free
mass), which affects the Dirac spinors used in the calculations of
the $NN$ scattering matrix. A linear dependence of the effective
mass of nucleons on the nuclear density was assumed in the
calculations.

The results of the polarization studies:

1. The difference of the final proton polarizations $P_1$ and $P_2$ found in
the PWIA, DWIA and DWIA* is quite small (less than 0.005) for all nuclei under investigation.

2. The difference between the PWIA and DWIA results is small. This indicates that
the distortion in the conventional non-relativistic framework does not
play any essential role in the polarization for the  kinematic
conditions under consideration (the transferred momenta $q$~=~3.2$\div$3.7~fm$^{-1}$).

3. Predictions of the DWIA* with relativistic corrections (distortion of
the proton Dirac spinor in nuclear medium) are close to experimental data for
the forward scattered proton polarization $P_1$.

4. A significant difference is observed between the measured polarization of
the scattered proton $P_1$ and that of the recoil proton $P_2$.

   Note that the difference between the measured polarizations $P_1$ and $P_2$ was also
observed in the reaction with the $D$--shell protons of the $^{28}$Si nucleus and
was not seen in the reaction with the $P$--shell protons of the $^{12}$C (Fig.~4).
\begin{figure}
\centering\epsfig{file=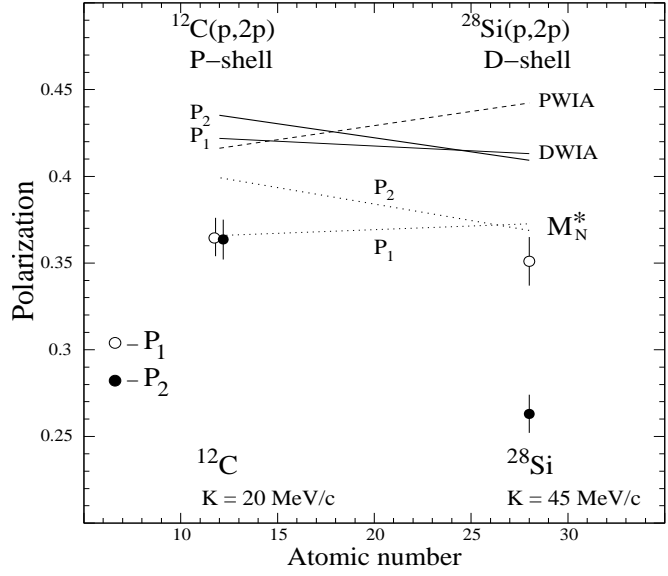,width=.6\textwidth,height=75mm}
\caption{\small Polarization in the  $(p,2p)$ reaction with the external shell protons
of the $^{12}$C  and $^{28}$Si.}
\end{figure}

The experimental data on the spin correlation parameters $C_{ij}$ in
the reactions with the $^4$He and
$^{12}$C are given in Fig.~5.
The dashed and dotted curves in the figure correspond to the PWIA
calculations of the $C_{nn}$ and $C_{s^,{s^{,,}}}$ parameters
using the current Arndt's group
phase-shift analysis (SP07). The mixed  $C_{s^,{s^{,,}}}$
parameter was found by taking into account its distortion in the
magnetic field of the spectrometers. The points at the $E_s$~=~0 correspond to the
elastic proton-proton scattering.

As seen in Fig.~5, the $C_{nn}$ data (as well as the $C_{s^,{s^{,,}}}$ data)
are described in the framework of the PWIA.
The question arises, there
is no the nuclear medium modification of the $C_{nn}$ parameter as it is
for the polarization of the final protons (Fig.~3)? Whether this fact is connected to
the strong
\begin{figure}
\centering\epsfig{file=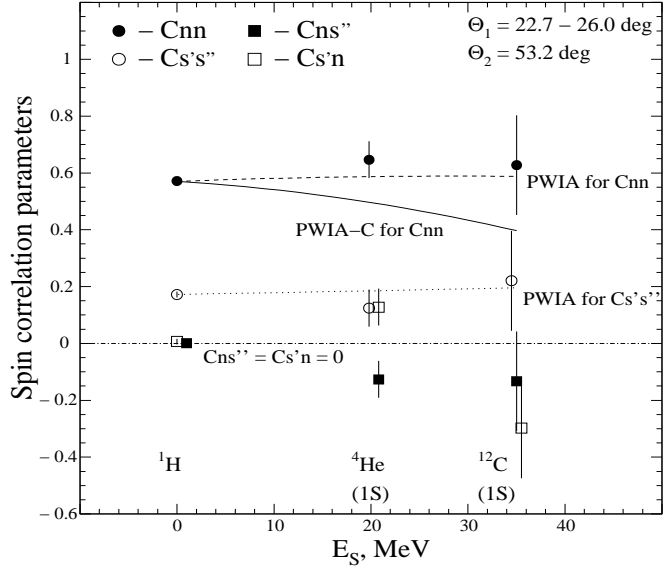,width=.6\textwidth,height=75mm}
\caption{\small Spin correlation parameters $C_{ij}$ in the $(p,2p)$ reaction at 1 GeV
with the $S$--shell protons of the $^4$He and $^{12}$C nuclei at the
secondary proton scattering angles $\Theta_2$~=~53.22$^\circ$,
$\Theta_1$~=~24.21$^\circ$ and $\Theta_2$~=~53.22$^\circ$, $\Theta_1$~=~22.71$^\circ$,
respectively. The points at $E_s$~=~0 correspond to the
elastic proton-proton scattering ($\Theta_1$~=~26.0$^\circ$,
$\Theta_{cm}$~=~62.25$^\circ$). The curves are the results of calculations
described in the text.}
\end{figure}
polarization dropping for the recoil proton?
It is possible that some spin-flip mechanism compensates the nuclear medium effect
in the $C_{nn}$.

Due to the parity conservation in the elastic proton-proton
scattering, the spin correlation parameters $C_{ns''}$ and
$C_{s'n}$ should be equal to 0. This is confirmed by the experimental
data at the $E_s$~=~0 in the Fig.~5. For the $(p,2p)$ reaction, we see some deviation of
the parameters from zero. It may be related to the spin-flip mechanism mentioned above.
Note that test calculations of all spin correlation parameters
for the accidental coincidence background give zero values as should be expected.

To find an explanation of the observed effects, let us assume that there is a
spin-flip interaction of the recoil (nuclear) proton with the residual nucleus, which is not
taken into account by the theoretical approaches. This additional interaction mechanism, governed
by the Pauli exclusion principle in a nucleus, reverses the proton spin direction
and, as a consequence, changes the signs of the polarization and the spin correlation parameter
$C_{nn}$.

The relative contribution ($\alpha$) of the spin--flip mechanism in the interaction with a residual nucleus,
which is mainly determined by the proton-nucleon re-scattering at small angles, can be found
from experiment via the relative polarization dropping ($g_p$) for the recoil proton. First define
the averaged polarization of the recoil proton:
\begin{equation}\label{equation:p2}
<{P}_2>~=~\frac{P_2+\alpha(-P_2)}{1+\alpha}~=~\frac{P_1+\alpha(-P_1)}{1+\alpha}~=~\frac{(1-\alpha)P_1}{1+\alpha}.
\end{equation}
In the equation we used the fact that all employed theories
give equal values of the polarizations $P_1$ and $P_2$.
The averaged value of the $C_{nn}$ can also be calculated using the equation:
\begin{equation}
<{C}_{nn}>~=~\frac{C_{nn}+\alpha(-C_{nn})}{1+\alpha}~=~\frac{(1-\alpha)C_{nn}}{1+\alpha}.
\end{equation}

 The relative polarization dropping $g_p$ is determined as:
\begin{equation}
g_p=\frac{P_1-<{P}_2>}{P_1}=g_{C_{nn}}=\frac{C_{nn}-<{C}_{nn}>}{C_{nn}}=\frac{2\alpha}{1+\alpha}.
\end{equation}
It can be seen that the proposed spin--flip interaction couples in simple form the relative
dropping of the polarization  and the $C_{nn}$ parameter $g_p$~=~$g_{C_{nn}}$.
From experimental data we find $g_p$($^4$He)~=~0.153$\pm$0.018, $g_p$($^{12}$C)~=~0.325$\pm$0.031
and make corrections to the PWIA calculations using the formula $C_{nn}$-cor~=~(1-$g_p$)$C_{nn}$ (the solid curve,
PWIA-C, in Fig.~5). One can see from the figure that the the experimental $C_{nn}$ points
lie above the curve. So it can be expected that the nuclear medium modification enhances the
$C_{nn}$ parameter, while the polarization is reduced.

From the experimental $g_p$ data, the probability of the spin--flip interaction can be
defined for the corresponding residual nuclei: $\alpha$($^3$H)~=~0.083$\pm$0.010,
$\alpha$($^{11}$B)~=~0.194$\pm$0.022.

What could be the nature of the considered spin--flip interaction?
It was first time proposed by D.I.~Blokhintsev that there are the fluctuations of nuclear
density in nuclei, or the dense nucleon associations [15].
The reflection of the recoil proton off the objects
is similar to the spin--flip interaction considered above. As a result, a proton belonging
to a correlation, with opposite spin direction (due to the Pauli principle) is
detected.

Nucleon correlations are intensively studied in the JLAB using electron beam.
The CLAS collaboration gives the probability for a given nucleon to belong to a two-nucleon
correlation in nucleus with A nucleons $a_{2N}$($^3$He)~=~0.080$\pm$0.016,
$a_{2N}$($^{12}$C)~=~0.193$\pm$0.041 [16].

 We can see that there is a coincidence between the PNPI $\alpha$
and the JLAB $a_{2N}$ for the corresponding residual nuclei. The model of the
spin--flip interaction for explanation of the PNPI polarization data is currently being developed.
Preliminary results suggest that the ratio of the $\alpha$ and $a_{2N}$
is very close to unity.\\
\newpage
{\bf\large References}\\
\\
\\
 1. G.E.~Braun and M.~Rho, Phys. Lett. \textbf{66}, 2720 (1991).\\
 2. C.J.~Horowitz and M.J.~Iqbal, Phys. Rev. \textbf{C 33}, 2059 (1986).\\
 3. G.~Krein {\it et al.}, Phys. Rev. \textbf{C 51}, 2646 (1995).\\
 4. O.V.~Miklukho {\it et al.}, Nucl. Phys. \textbf{A 683}, 145 (2001).\\
 5. V.A.~Andreev {\it et al.}, Phys. Rev. \textbf{C 69}, 024604 (2004).\\
 6. O.V.~Miklukho {\it et al.}, Phys. Atom. Nucl. \textbf{69}, 474 (2006).\\
 7. G.C.~Hillhouse and T.~Noro, Phys. Rev. \textbf{C 74}, 064608 (2006).\\
 8. O.V.~Miklukho {\it et al.}, Phys. Atom. Nucl. \textbf{73}, 927 (2010).\\
 9. G.~Jacob {\it et al.}, Nucl. Phys. \textbf{A 257}, 517 (1976).\\
10. K.~Hatanaka {\it et al.}, Phys. Rev. Lett. \textbf{78}, 1014 (1997).\\
11. T.~Noro {\it et al.}, Nucl. Phys. \textbf{B 633}, 517 (2000).\\
12. N.S.~Chant and P.G.~Roos, Phys. Rev. \textbf{C 27}, 1060 (1983).\\
13. W.O.~Lock and D.F.~Measday, {\it Intermediate-Energy Nuclear Physics}\\
    (Methuen, London, 1970; Atomizdat, Moscow, 1973).\\
14. O.V.~Miklukho {\it et al.}, Phys. Atom. Nucl. \textbf{76}, 871 (2013).\\
15. D.I.~Blokhintsev, Sov.J. ZhETF \textbf{33}, 1295 (1957) [in Russian].\\
16. K.S.~Egiyan {\it et al.}, Phys. Rev. Lett. \textbf{96}, 082501 (2006).\\

\end{document}